\documentclass[10pt,twocolumn]{article} 
\pdfoutput=1

\usepackage[utf8]{inputenc} 
\usepackage[letterpaper,margin=1in]{geometry} 
\usepackage{graphicx} 
\usepackage{color,amsmath}
\usepackage{booktabs}
\usepackage{dblfloatfix}
\usepackage{supertabular}
\usepackage{multirow, array}

\makeatletter
\def\bstctlcite#1{\@bsphack
\@for\@citeb:=#1\do{%
\edef\@citeb{\expandafter\@firstofone\@citeb}%
\if@filesw\immediate\write\@auxout{\string\citat
ion{\@citeb}}\fi}%
\@esphack}
\makeatother

\usepackage[utf8]{inputenc}

\title{Description of the EMOD-HIV Model $v0.7$}
\author{Anna Bershteyn*, Daniel J.~Klein*, Edward Wenger, Philip A.~Eckhoff\\\\
	Intellectual Ventures Laboratory, $1555$ $132^{nd}$ Ave.~NE, Bellevue, WA $98005$\\
*A. Bershteyn and D.J. Klein contributed equally to this work\\
Correspondence: abershteyn@intven.com}
\date{} 

\begin{document}
\maketitle

\begin{abstract}
The expansion of tools against HIV transmission has brought increased interest in epidemiological models that can predict the impact of these interventions. The EMOD-HIV model was recently compared to eleven other independently developed mathematical models of HIV transmission to determine the extent to which they agree about the potential impact of expanded use of antiretroviral therapy in South Africa. Here we describe in detail the modeling methodology used to produce the results in this comparison, which we term EMOD-HIV $v0.7$. We include a discussion of the structure and a full list of model parameters. We also discuss the architecture of the model, and its potential utility in comparing structural assumptions within a single modeling framework. 
\end{abstract}

\section{Introduction}

This document provides a detailed description of the ``EMOD-HIV'' population model of HIV that was compared to several other HIV models in a recent article by Eaton et al.~published in PLoS Medicine \cite{eaton_systematic_2012}.  The EMOD-HIV model was designed by researchers in the Epidemiological Modeling group at Intellectual Ventures Laboratory as part of a larger effort to enable broad accessibility of modeling and quantitative analysis tools for planning and execution of infectious disease control. The group has thus far developed mathematical models of malaria \cite{eckhoff_malaria_2011}, polio, and tuberculosis in addition to HIV. The models and their accompanying software infrastructure are designed primarily for use by disease modelers, researchers, epidemiologists, and public health professionals seeking to simulate infectious disease conditions and evaluate the effectiveness of different eradication or mitigation approaches. 

This document describes only the HIV transmission model, and not the other diseases, parallel computing infrastructure, or additional features of the software not utilized in the Eaton $2012$ model comparison.

\begin{figure*}[t]
\centering
\includegraphics[width=.9\textwidth]{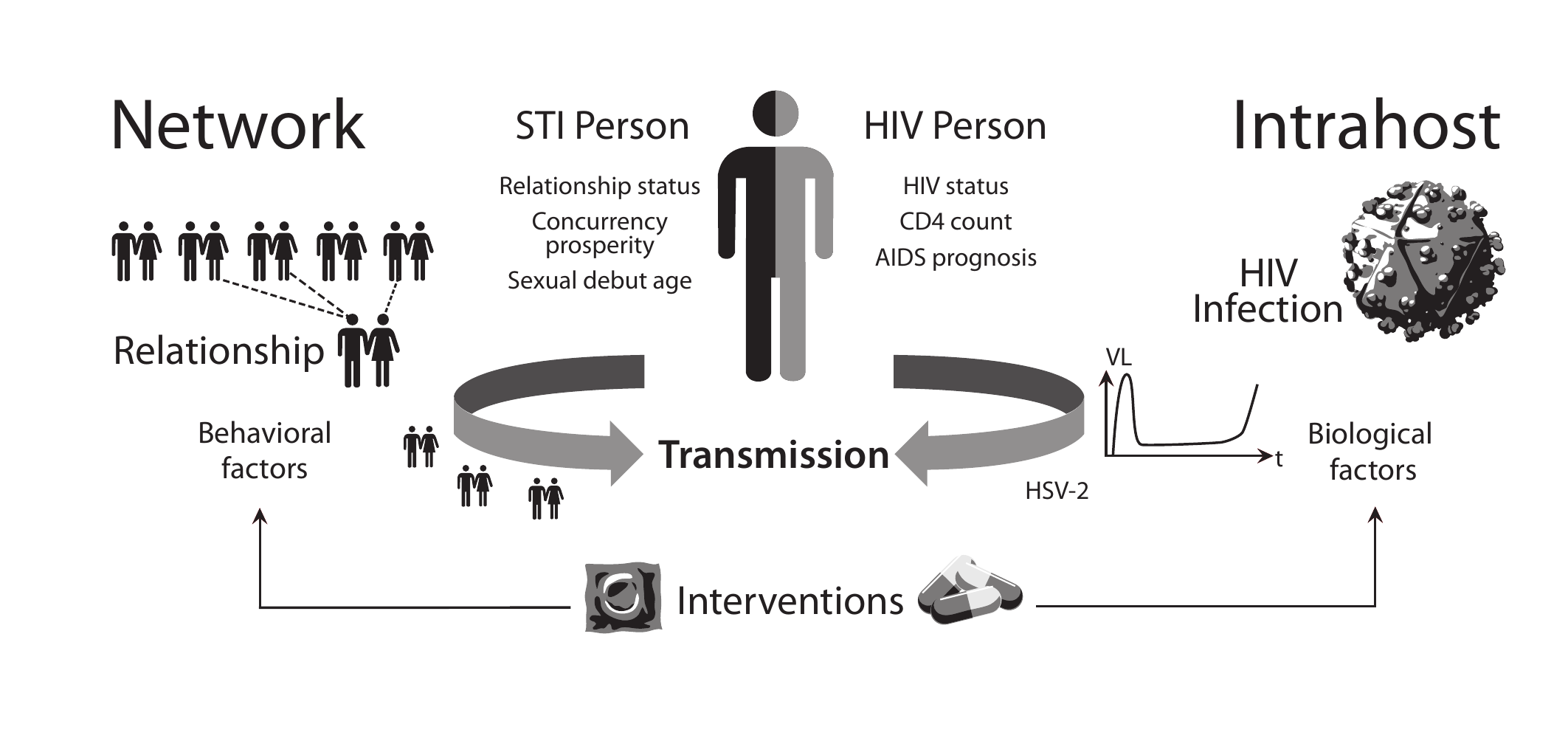}
\caption{The EMOD-HIV model is designed to cleanly separate components related to the contact network (Network) from within-host biology of HIV infection (Intrahost). Both of these build upon a generic “Person” whose properties are more broadly applicable in epidemiological models (e.g., age, gender, and vital statistics). This Person object is specialized into ``STI Person,'' containing attributes related to behavior within the contact network and potentially applicable to other STIs, and a further specialized ``HIV Person'' object containing attributes that are specific to the biology of HIV.  Interventions can modify factors in both modules that influence transmission, such as the rates of condom use or use of antiretroviral therapy.  In $v0.7$, each HIV-discordant interaction is modeled explicitly, and the transmission probability is calculated based on behavioral factors (type of sexual contact, condom use, etc.) and biological factors (stage of HIV disease, co-infections, ART, etc.) after interventions are applied.}
\label{f:structure}
\end{figure*}

The Eaton $2012$ model results from EMOD were generated from a preliminary version of the model, which we will denote $v0.7$.  While $v0.7$ includes all the basic elements of a HIV simulation, to be discussed shortly, it lacks some elements that could have added heterogeneity to behavior and biology of HIV transmission, such as variable viral loads and disease stage durations. Additionally, the model parameters were calibrated using a rudimentary search algorithm that did not allow for parameter uncertainty to be incorporated in error bound calculations.  More detailed and systematically calibrated versions of the model are in progress. Outputs from $v0.7$ should be considered preliminary.

\section{Overview of the Model}

EMOD-HIV $v0.7$ is an individual-based stochastic simulation of sexual and vertical HIV transmission implemented in C++ and parameterized for the setting of South Africa.  More specifically, $v0.7$ targets South Africa at a national level, with sexual mixing parameters based on data from KwaZulu-Natal province.  

Geographic distribution of the population was not explicitly modeled in $v0.7$, but an explicit network of human relationships was formed based on pair formation rules, and allowed to evolve over time. Though capable of modeling homosexual partnerships and commercial partnerships (in which providers are distinguished from patrons), $v0.7$ utilized only heterosexual relationships due to their importance in the modern sub-Saharan Africa transmission setting \cite{shisana_nelson_2002}. Thus, HIV transmission could only result from heterosexual intercourse or mother-to-child vertical transmission in $v0.7$. Three types of heterosexual relationships were modeled, each characterized by different distributions for ages of entrants, durations, and levels of concurrency. The youngest, shortest-duration relationship type was most likely to permit the individuals to simultaneously seek additional relationships. Thus, although the propensity to form multiple simultaneous partnerships was not explicitly age-dependent, the types of partnerships preferred by younger individuals implicitly increase their likelihood to enter concurrent partnerships.

The remainder of this article describes the EMOD-HIV $v0.7$ model in detail, with subsections for demographics, individual properties, the pair formation algorithm, relationship flow, sexual relationships, sexual acts, HIV transmission, HIV infection (with and without treatment), and the antiretroviral therapy (ART) intervention.  The article concludes with a brief discussion of future work.

\section{Detailed Model Description}

\subsection{Model Structure}
The EMOD-HIV model propagates forward in time using a combination of discrete events and time steps.  Discrete events, implemented using C++11 lambda expressions, allow for one or more function calls to be scheduled for execution at a future time in the simulation.  

As with all the EMOD disease models, the HIV model utilizes object-oriented programming design principles such as interfaces, factories, and observers to achieve a highly modular architecture, thus enabling comparisons of structural assumptions and different levels of model detail. For example, parameters specific to the within-host biology of disease progression and to behavioral propensities that drive pair formation have been separated into different inheritance classes. Thus, modules or sub-modules can be interchanged while leaving other portions of the model intact. Figure \ref{f:structure} provides diagram of the architecture, the components of which will be discussed in detail in the sections that follow.

\subsection{Demographics}
Vital dynamics within the EMOD-HIV model are derived from fertility and mortality tables that are passed to the model as input.  In $v0.7$, these values came from the Actuarial Society of South Africa (ASSA) AIDS and Demographic Model \cite{actuarial_society_of_south_africa_assa_2005}, using linear interpolation between age-specific data points from ASSA to construct a cumulative probability distribution function (CDF) of death date from a person’s date of birth. Upon instantiation, individual agents in the model sample stochastically from this CDF using an inverse transform of this distribution. Female agents similarly sample the age at next childbirth, if any, upon instantiation and birth of a previous child. Pregnancy is not linked to relationship status in $v0.7$, although newly born individuals are linked to a mother. The fertility rate changes by simulation year and female age, and the available estimates range from 1980 to 2020. Values outside of this range were chosen by ``clamping'' - i.e., choosing the nearest value within the range.  Clamping was also used when necessary to determine the non-AIDS mortality rate, which varies by gender, age, and simulation year.

Mother-to-child transmission (MTCT) is enabled in $v0.7$ using the linkage between the mother and offspring created by the demographics module.  The baseline rate of MTCT is $1$ in $3$ births \cite{msellati1995rates,newell_child_2004-1,orlando_cost-effectiveness_2010}. Antiretroviral therapy (ART) reduces the transmission rate to $1$ in $100$ births \cite{chibwesha_optimal_2011}, except in the final 9 months prior to death due to ART failure, when it is assumed to have no effect on MTCT because HIV-related death after ART initiation is assumed to occur due to virological failure. (As we will show, many individuals do not reach the time of virological failure and death on ART because non-HIV mortality occurs at an earlier time - i.e., they achieve a normal lifespan while on ART.) Additionally, $45\%$ of mothers not on ART receive single-dose Nevirapine prophylaxis (sdNVP), which reduces the MTCT rate to $1$ in $10$ births \cite{jackson2003intrapartum, guay_intrapartum_1999, martinson_transmission_2007}.

\subsection{Individual properties}

Individuals within the simulation have a variety of properties, represented by continuous or discrete state variables.  Some properties are static throughout life, and others dynamically change through the course of the simulation either in response to aging or simulation events such as infection.  Included in these properties are complex dynamic objects for HIV infection and relationships, to be described shortly.  Static properties are assigned upon instantiation (simulation initialization or birth after the beginning of the simulation) and include gender, time of birth, time of non-HIV death, STI co-infection, and male circumcision. In $v0.7$, the latter two are sampled randomly at birth with respective probabilities of $7\%$ and $42\%$ \cite{shisana_nelson_2002}. Dynamic properties include HIV status, history of ART program participation and drop-out if applicable, and relationship participation. 

\subsection{The Pair Forming Algorithm}

Individuals enter a sexually active state at an age randomly sampled from a Weibull distribution with a shape parameter of $20$ and a mean age of $16.5$ years and $15.5$ years for males and females, respectively, and never permitting sexual activity earlier than $13$ years of age \cite{shisana_nelson_2002,pettifor_young_2005,mcgrath2009age,shisana_south_2008}. These values were determined by calibration, using national survey results as a prior \cite{shisana_nelson_2002,pettifor_young_2005,shisana_south_2008,shisana_south_2010,johnson_second_2010}. Individuals that have exceeded their stochastic sexual debut age are able to enter into three types of sexual relationships: transitory, informal, and marital.  Three Boolean static properties of individuals, termed ``extra-relational flags,'' determine whether an individual can enter additional concurrent relationships while participating in a relationship of each of the three types. In $v0.7$, these flags are stochastically assigned at instantiation with probabilities of $80\%$, $30\%$, and $10\%$ for transitory, informal, and marital relationships, respectively. Note that these flags indicate the ability of an individual to participate in concurrent relationships, and were determined by calibration, but the resulting point prevalence of concurrency ends up close to values found in \cite{pettifor_young_2005}. These extra-relational flags create heterogeneities in HIV risk: an individual who receives all three flags, for example, would be likely to participate in many more lifetime partnerships than one who receives no flags.

The algorithm used to take individuals seeking relationships and form pairs is called the pair formation algorithm (PFA).  Each of the three relationship types uses the same PFA algorithm, but operates on different data about the age distribution and age gaps within partnerships. 
 
\begin{figure*}[t]
\includegraphics[width=\textwidth]{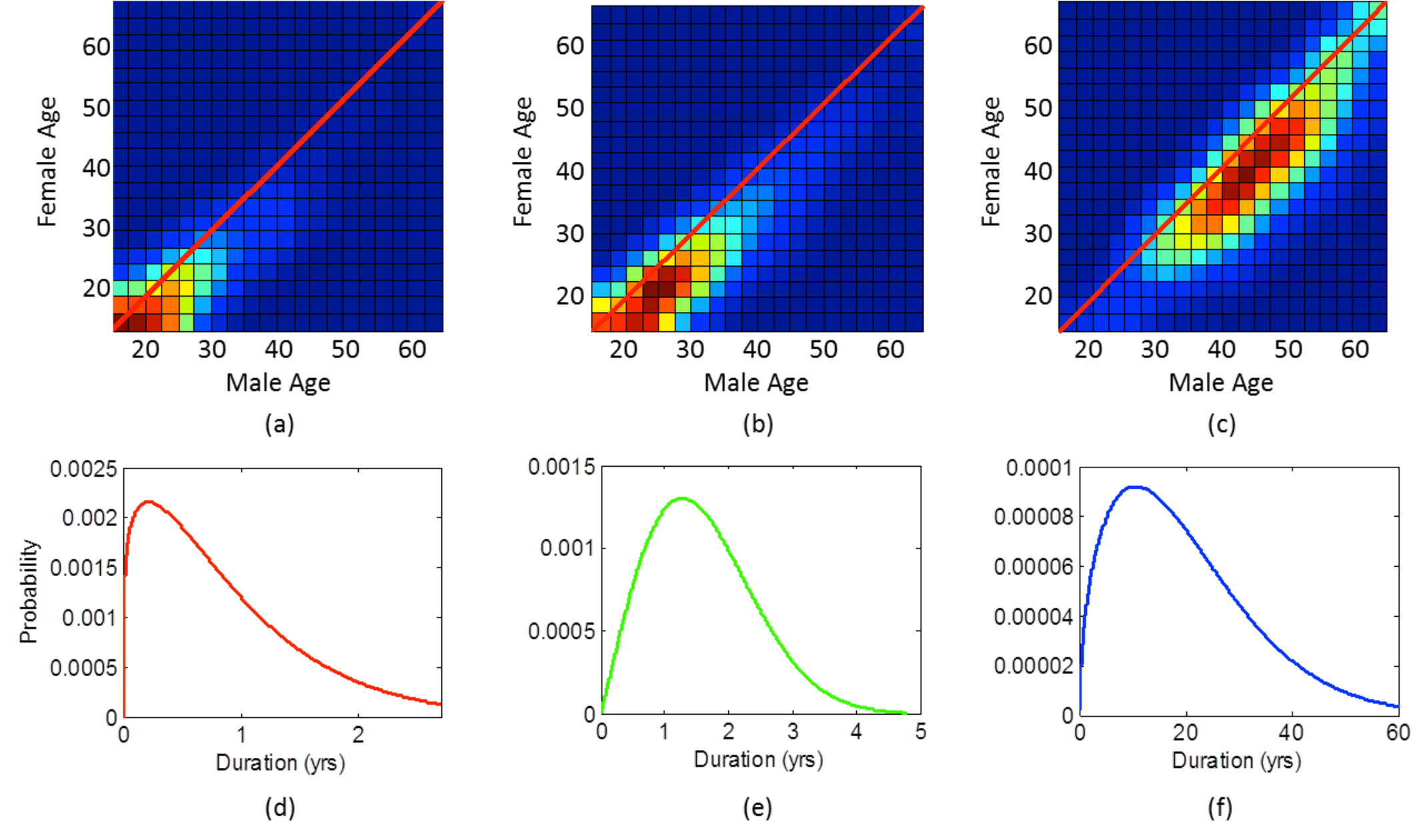}
\caption{Transitory (left column), informal (middle column), and marital (right column) relationships are characterized by their matrix of partnership age distributions (MOPAD, top row) and by their Weibull-distributed duration (bottom row; note the difference in scale on the time axis). On average, transitory relationships recruit the youngest individuals, create the smallest age gap between partners, and have the shortest durations, while marital partnerships recruit the oldest individuals, create the largest age gap between partners (approximately 5 years), and have the longest durations.}
\label{f:relationships}
\end{figure*}

PFA takes as input a matrix, called the matrix of partnership age distributions (MOPAD), in which the entry at position $(i,j)$ is the probability of a male of age bin $i$ pairing with a female of age bin $j$.  The three MOPADs used in the can be seen in Figure \ref{f:relationships}(a-c) and are based on published data by Ott et al.~from the Africa Centre for Health and Population Studies \cite{ott_age_gaps_2011}. Ott et al.~report their data in $5$-year age bins; these were sub-divided to $2.5$-year age bins, shifted to represent incident rather than prevalent relationships, and smoothed using a Gaussian kernel.  Membership in an age bin was determined based on each individual's time of birth.

\begin{figure}[t]
\includegraphics[width=\columnwidth]{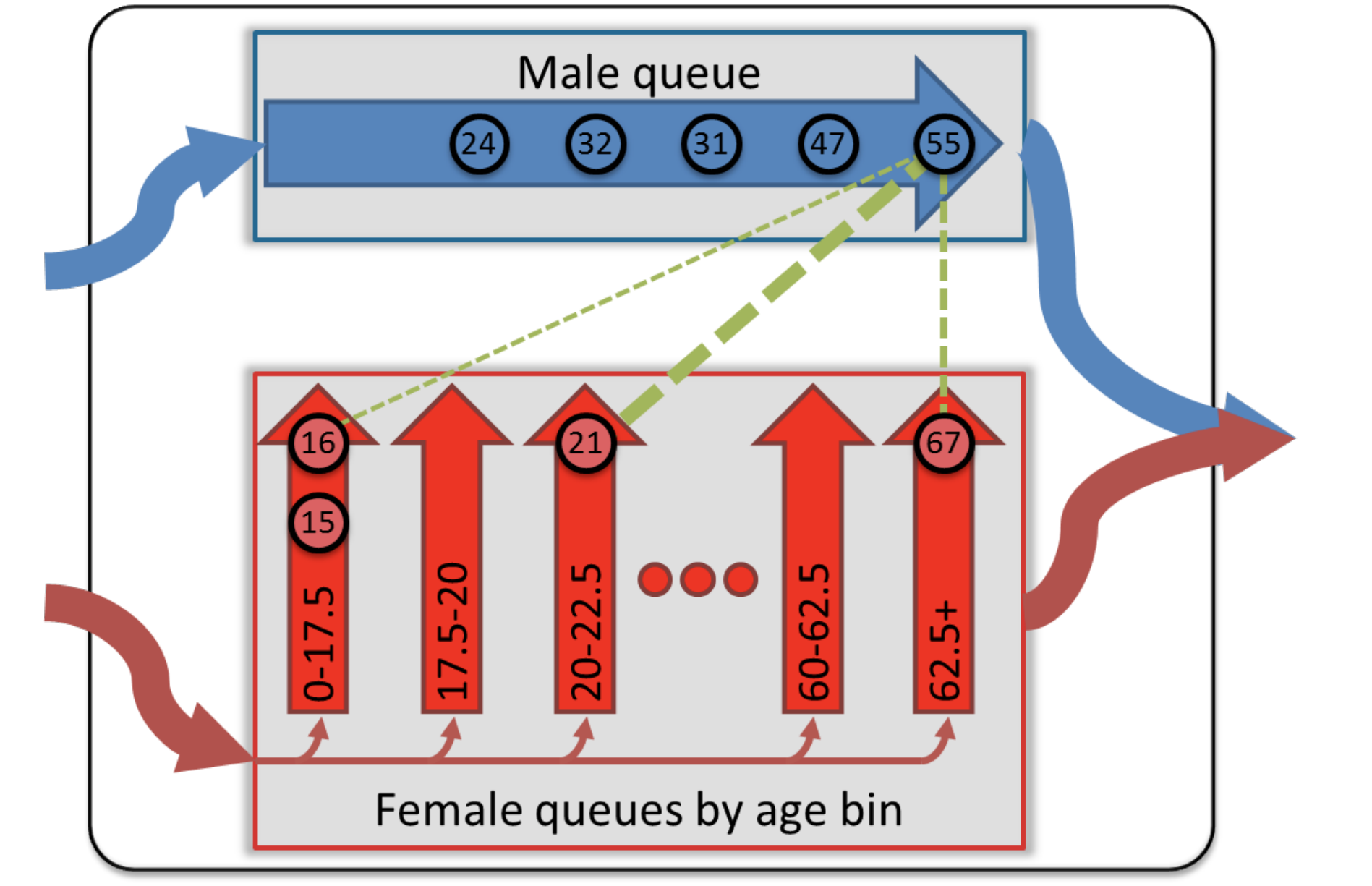}
\caption{A diagram of the Pair Formation Algorithm (PFA).  Males and females enter from the left. Males form a single queue, while females are sorted into queues according to their age, represented above as labeled arrows. Males and females accumulate in the order they entered and remain in their queues until the PFA is processed. Processing takes place once per day in the transitory and informal PFAs, and once every 14 days in the marital PFA. During processing, the individual at the head of the male queue selects a partner based on his MOPAD-derived conditional probability weighting (represented by dashed green lines).  The paired individuals exit the algorithm to the right. Only individuals with no available partners of their chosen type are left in queues after processing is complete.}
\label{f:PFA}
\end{figure}

Individuals enter the PFA seeking a relationship, and are immediately placed into a gender-specific queue, see Figure \ref{f:PFA}.  There is a single queue for all males, and for females, there are separate queues for each $2.5$-year age bin.  The queues fill over a period of time (fourteen days for marriage and one day for transitory and informal), after which relationships are formed by a processing step.  The processing step works linearly through the male queue, starting from individual at the head of the queue.  A female partner for the head male is determined based on age and availability.  More specifically, a male of age bin $m$ samples a partner age bin, $f$, from 
\begin{equation}
p(F=f) \propto \begin{cases}
p(F=f|M=m)&\text{ if $N_f>0$}\\
0&\text{ otherwise}.
\end{cases}
\end{equation}
Here, $N_f$ is the number of females queued in age bin $f$ and $p(F=f|M=m)$ is the conditional distribution derived from the MOPAD.  If $p(F=f)$ is identically zero, the male remains in the queue for the next processing round.  Otherwise, the female of the sampled age bin who has been waiting the longest is selected as the partner.  The paired individuals are removed from their respective queues, and the process is repeated. 

\subsection{Relationship Flow}

The pair formation algorithm can produce relationships distributed according to the MOPAD matrix only if the males and females entering the algorithm have age distributions that match the respective marginals of the MOPAD, and the total number of females matches the total number of males.  The (exponential) rates at which individuals in each age bin and gender enter for each relationship type are adjusted daily to form the relationships dictated by the MOPAD. The MOPAD dictates only the relative number of relationships formed between pairs of different ages, but not the absolute number of relationships formed by the PFA.  This total throughput of relationships formed is set such that the expected number of individuals seeking a relationship of a particular type, after rate adjustment to meet the MOPAD, matches the number of males and females that would have sought that relationship had the rates not been adjusted for the MOPAD. These unadjusted rates are listed in Table \ref{t:relationships}.  Thus, the MOPAD adjustment changes the age distributions of the individuals seeking relationships, but not the total number of each type of relationship formed.

\begin{table}[t]
\caption{Base exponential relationship formation rates and Weibull dissolution distribution parameters.}
\centering
\begin{tabular}{cccc}
\toprule
&&\multicolumn{2}{c}{Duration}\\\cmidrule(r){3-4} 
	&{\bf Rate}&	{\bf Mean}&	{\bf Shape}\\
	&(per year)&	(per year)&	\\\hline
Transitory&	1/2&	1/0.9	&1.2\\
Informal&	1/1.9&	1/1.8&	2.0\\
Marital&		1/30&	1/20&	1.5\\
\bottomrule
\end{tabular}
\label{t:relationships}
\end{table}

This adaptive daily rate control allows the model to automatically discover the rates of relationship entry that are consistent with the MOPAD. It is conceivable, however, that events causing large demographic shifts might change the MOPAD. For example, when comparing simulations with universal HIV treatment versus no treatment, it is conceivable that demographic influence of the disparate AIDS death tolls should cause the MOPADs to diverge. Therefore, adaptive rate control is disabled after an initial burn-in period of $15$ years, after which the entrance rates remain fixed at their final controlled values and the MOPAD is permitted to change as the simulation progresses.

The calculation of the total relationship formation rate (Table \ref{t:relationships}) and adjustment of individual relationship entry rates are based only on individuals eligible to form a new relationship. Eligibility to form a new relationship depends on several factors.  Individuals that have not surpassed their debut age are not eligible for any relationships, and all individuals must be at least $15$ years old to marry.  Post-debut individuals who are not already in a relationship are eligible to add any of the three relationships.  Young individuals (e.g.~$<20$) will tend to enter a transitory relationship and older individuals (e.g.~$>40$) will tend to enter a marital relationship due to their different age distributions shown in Figure \ref{f:structure}.  

\begin{table*}[ht]
\caption{Sigmoidal condom ramp parameters.}
\centering
\begin{tabular}{ccccc}
\toprule
&{\bf Low ($l$)}&	{\bf High} ($h$)&	{\bf Middle Year} ($t_0$)&	{\bf Rate} ($r$)\\
Transitory&	0.0125&	0.6375&	1999&	0.3\\
Informal&	0.0125&	0.5125&	1999&	0.25\\
Marital&	0.000625&	0.200625&	2002&	0.225\\
\bottomrule
\end{tabular}
\label{t:condomRamp}
\end{table*}

Once in a relationship, the previously described Boolean concurrency flags determine if the individuals is allowed to add additional (concurrent) relationships, after a waiting period of $60$ days.  For example, if the transitory concurrency flag is true, the individual is allowed to add additional relationships of any type as long as their only ongoing relationships are of transitory type. Coital dilution, discussed later, is applied to any concurrent partnerships, and the maximum number of simultaneous relationships restricted to three transitory, two informal, and one marital relationship(s). Note that, under these assumptions, a small subset of the population will have all three flags true, and comprises an especially high-risk subgroup. Note also that in this model, the formation of a marital partnership, which has the lowest probability of permitting concurrency and the longest duration, will often ``protect'' individuals from taking on additional partnerships later in life.

\subsection{Sexual Relationships}

Sexual relationships are instantiated by the three PFAs and have characteristics including the participating individuals, condom usage rate, formation time, and scheduled breakup time.  When the relationship forms, its breakup time is sampled from a PFA-type-dependent Weibull distribution, see Figure \ref{f:relationships}(d-f).  The mean and shape parameters of these distributions are listed in Table \ref{t:relationships}.  Note that relationships can terminate early if either partner dies, and the surviving partner is treated as eligible, equivalently to a someone who has become eligible through a relationship breakup.

Each sexual relationship has a fixed condom usage probability, called the base rate, which is determined at the time the relationship is created.  The functional form of the base condom usage probability is an increasing sigmoid, 
\begin{equation}
P(t)=  \frac{h}{1+e^{-R(t-t_0)} }+l,
\end{equation}
the parameters of which depend on the type of relationship (see Table \ref{t:condomRamp}).  The relationship type-dependence of these parameters is to account for the fact that condom usage is lower in longer-term relationships, while the time-dependence accounts for reported increases in condom usage over time.  Self-reported condom usage probabilities vary greatly.  The values in Table \ref{t:condomRamp} come from calibrating parameters that were initially determined from data given in the $2002$, $2005$, and $2008$ HSRC national surveys \cite{shisana_nelson_2002,shisana_south_2008,shisana_south_2010}.

For half of the relationships with a base rate above $10\%$ and below $90\%$, a modified rate of either $10\%$ or $90\%$ was used to add additional heterogeneity by increasing the variance.  This was implemented in a way that ensured that the expected condom usage probability remained equal to the base rate.

\subsection{Coital Acts}

Individual coital acts are simulated for HIV-discordant relationships only.  When an individual becomes infected, the discordancy states of all relationships involving the individual are updated.  Within HIV-discordant relationships, the base time between coital acts is exponentially distributed with a mean of three days \cite{wawer_rates_2005}.  For individuals in two, three, or four-plus concurrent partnerships, our model additionally accounts for coital dilution, i.e., a negative correlation between the number of simultaneous partnerships and the frequency of coital acts within these partnerships \cite{sawers_hiv_2011,sawers_concurrent_2010}. While some studies suggest a very strong effect of coital dilution \cite{blower_sex_1993,reniers_polygyny_2010}, others find little evidence for this phenomenon \cite{josephson_does_2002,pebley_polygyny_1988}. We therefore chose a relatively modest parameterization of coital dilution, in which the act frequency per relationship for an individual belonging to two, three, or four (or more) concurrent relationships is multiplied by $0.75$, $0.60$, and $0.45$, respectively.  Condom usage is assigned to each coital act based on the relationship's usage probability, and higher-risk anal intercourse is selected $2\%$ of the time.

\subsection{Stages of HIV Infection}

We assume three stages of untreated HIV infection: acute, latent, and AIDS.  In $v0.7$, the durations of the acute and AIDS stages are $2.9$ months and $9$ months, respectively, for all individuals \cite{hollingsworth_hiv_1_2008}. The total survival time with HIV is sampled stochastically based on the individual's age at the time of infection (discussed below), and the duration of the latent phase is the total survival time minus the duration of the acute and AIDS stages. If the total survival time is shorter than $11.9$ months, then the AIDS phase is shortened, and in rare cases when survival is shorter than $2.9$ months, the remaining acute phase is shortened to fit within the survival duration.

\subsection{HIV Transmission}

The probability of HIV transmission within a discordant relationship is calculated on a per-act basis and varies by disease stage. The per-act transmission probabilities for the acute, latent, and AIDS stage are $3.4E-2$, $1.3E-3$, and $9.4E-3$, respectively.  The relative magnitude of these values in each stage of disease comes from the per-time transmission rates measured in HIV-discordant couples in Uganda \cite{hollingsworth_hiv_1_2008}.  These probabilities have been multiplied by the inverse act frequency and increased by $50\%$ as a result of the model calibration process.  Note that these transmission probabilities apply in the absence of any interventions or other cofactors.  The per-act transmission probability is multiplied by applicable act-dependent factors, which are described in Table \ref{t:cofactors}.  There is no gender bias in transmission other than the male circumcision cofactor, which reduces the circumcised male's HIV acquisition probability by $60\%$ \cite{auvert_randomized_2005, bailey_male_2007, gray_male_2007}.

\begin{table*}[ht]
\caption{Cofactors used in determining the HIV transmission probability.}
\centering
\begin{tabular}{cc}
\toprule
{\bf Cofactor}	&{\bf Transmission Probability Multiplier}\\
Male circumcised (F$\to$M transmission only)&	0.4\\
STI present&	4.0\\
Condom used&	0.1\\
Anal intercourse (infected receptive female partner)&	4.0\\
Anal intercourse (infected insertive male partner)&	10\\
\bottomrule
\end{tabular}
\label{t:cofactors}
\end{table*}
\begin{figure*}[t]
\centering
\includegraphics[width=0.8\textwidth]{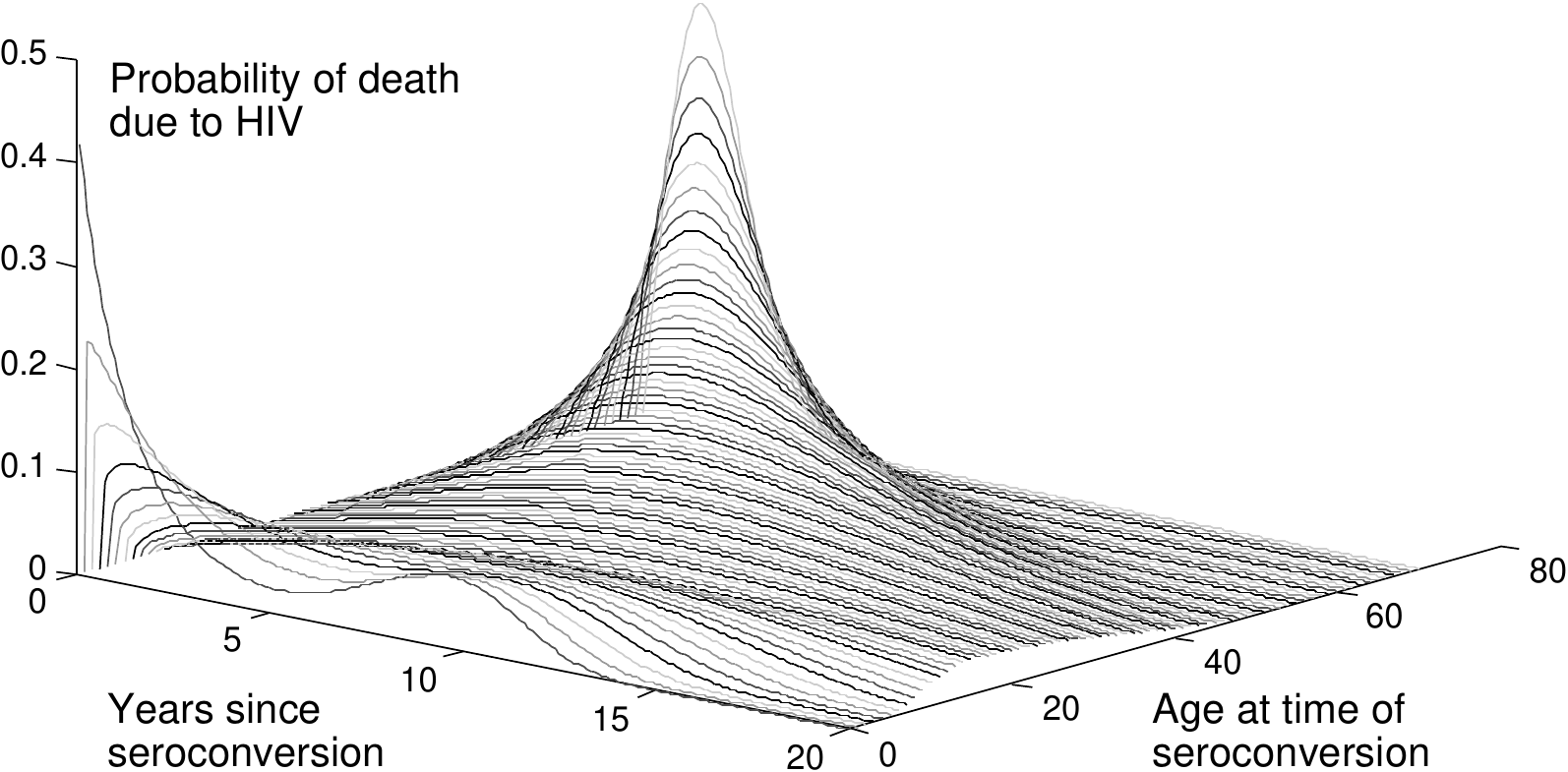}
\caption{The untreated HIV prognosis of an individual is determined at the time of infection using this age-dependent distribution. Before age $15$, a sum of two weighted Weibull distributions is used to represent the typically bimodal survival time of children. Beyond age $15$, survival time is Weibull-distributed, with older individuals having a shorter average survival time. Note that untreated HIV-infected individuals die at the end of their HIV prognosis or their normal lifespan (determined by the demographics module), whichever occurs earlier.}
\label{f:prognosis}
\end{figure*}

If the HIV positive individual is on ART and not experiencing treatment failure, then the transmission probability per act is set to $4\%$ of the latent-stage transmission probability \cite{cohen_prevention_2011}.  When failing ART, the transmission probability is set to the AIDS transmission probability.

\subsection{Survival with Untreated HIV Infection}
\label{s:survival_with_untreated_HIV_infection}

HIV prognosis is both heterogeneous and highly age-dependent \cite{todd_time_2007,lutalo_survival_2007,babiker_age_2001}, and is therefore sampled from an age-dependent distribution, shown in Figure \ref{f:prognosis}.  For individuals over the age of $15$, the distribution is Weibull-distributed with shape parameter $\kappa$ set to $2$ and an age-variable scale parameter $\lambda$ varying with age $a$ (in years) as follows \cite{todd_time_2007}:
\begin{equation}
\lambda = 21.182-0.2717a
\end{equation}

For those under $15$, the distribution is a convex combination of two Weibull distributions,
\begin{align}
\lambda_1&= 1.515+1.039a\\
\kappa_1&= 0.97+0.0687a\\
\lambda_2&= 10.0+0.474a\\
\kappa_2&= 5.39-0.226a,
\end{align}
distinguishing infant mortality and later child mortality~\cite{marston_estimating_2005_1}. The first distribution is used $65\%$ of the time.

CD4+ T-cell count is time-dependent based on the fraction remaining of the individual's prognosis $f$, calculated by dividing the time since infection by the total HIV prognosis.  It drops to $594$ cells/$\mu$L immediately upon infection to represent very rapid CD4 T-cell depletion (and partial rebound) during the acute phase (the detailed timecourse of which is not modeled here), and decreases down to $59$ cells/$\mu$L at death as follows:
\begin{equation}
\mathrm{CD4}(\mathrm{cells}/\mu \mathrm{L}) = (24.363 - 16.672 f)^2.
\end{equation}


\subsection{HIV Infection (with ART)}

Mortality for individuals receiving ART is determined using a multivariate Cox proportional hazards model from the International Databases to Evaluate AIDS - South Africa \cite{egger_cohort_2011,may_prognosis_2010}, which takes as input the individual's age, gender CD4 count, and body weight to estimate the prognosis. Note that this database utilizes ecological studies of ART efficacy, and thus includes ``real-world'' factors such as poor adherence, co-morbidities, misuse of ART, undetected drug resistance, etc., which may contribute to mortality on ART and not be captured in randomized controlled trials. 

In addition to the individual's age and gender, the WHO stage and body weight are estimated by assuming individuals progress from the pre-infection state through WHO stages from one to four stochastically, with the interval from WHO stage $1$ to $2$, $2$ to $3$, and $3$ to $4$ samples from a Weibull distribution with shape parameters $0.26596$, $0.19729$, and $0.34721$, respectively, and scale parameters $0.96604$, $0.99160$, $0.93560$, respectively \cite{johnson_modelling_2006}. Any remaining pre-treatment survival time is spent in WHO stage 4. Body weight is assumed to decline according to the midrange values of the WHO AIDS staging guidelines: from a pre-infection weight of $65$ kg to $62.1$ kg in Stage $1$, $57$ kg in Stage $2$, $50$ kg in Stage $3$, and $40.1$ kg in Stage 4. Calculation of CD4 decline was described earlier (see Section \ref{s:survival_with_untreated_HIV_infection}).

The survival time, $s$, for an individual on ART was calculated as \cite{may_prognosis_2010}:
\begin{equation}
s(\mathrm{years}) = \left( \frac{123.83}{m^{2.9} } \right) \log\left(\frac{1}{r}\right)^{2.9},
\end{equation}
in which $r$ is a uniformly distributed random number between $0$ and $1$, and $m$ is the appropriate multiplier applied for the individual based on the conditions listed in Table \ref{t:survival}. Mortality while on ART is assumed to be associated with a period of virological failure with the same duration and infectivity as the late-stage AIDS phase of untreated disease. For a majority of individuals, this period is longer than their HIV-independent survival, such that they never survive to realize the higher-infectivity phase at the end of successful viremic suppression by ART.

\begin{table}[t]
\centering
\begin{tabular}{cc}
\toprule
{\bf Condition}&	{\bf Multiplier}\\
WHO Stage 3 or 4&	2.7142\\
Age $>$ 40 years&	1.4309\\
Female gender&	0.6775\\\\
\multicolumn{2}{l}{CD4 at enrollment}\\ 
\cmidrule{1-1}   25-49&	0.7497\\
  50-99&	0.4258\\
  100-199&	0.3068\\
  200+&	0.2563\\
\bottomrule
\end{tabular}
\caption{Multipliers used in survival time calculation.}
\label{t:survival}
\end{table}

In the event of dropout from the ART program, an individual's recovery while on ART is represented by an increase in their CD4 count, which occurs independently of their starting CD4 count and saturates after three years on ART \cite{kaufmann_cd4_2003,binquet_modeling_2001,de_beaudrap_modeling_2009,li_long_lasting_1998,williams_hiv_2006}. The increase is calculated as \cite{kaufmann_cd4_2003}:
\begin{equation}
\text{CD4 increase}(\mathrm{cells}/\mu \mathrm{L})= 15.584 t  - 0.2113 t^2,
\end{equation}
in which $t$ represents the time, in months, that the individual has received ART. The increase saturates after $36$ months and does not allow the individual to exceed their pre-infection CD4 count. CD4 reconstitution is used to compute the untreated prognosis following dropout from the ART program, essentially turning back the clock on the infection to the fraction of remaining prognosis that corresponds to this CD4 count.

When an individual is re-enrolls in an ART program after a prior dropout, the process to re-compute the ART prognosis is identical to the initial enrollment; however, the individual's prognosis parameters may change as the individual will have aged in the intervening time. 

\subsection{ART Intervention}

The ART intervention was prescribed by the Modelling Consortium organizers \cite{eaton_systematic_2012}.   CD4 thresholds for treatment eligibility are $200$ cells/$\mu$L, $350$ cells/$\mu$L, and immediate.  Once eligible, an individual samples an enrollment date from an exponential distribution with one-year mean.  A fraction of the population is not allowed to enroll, as prescribed in the guidelines.  ART is not distributed preferentially by age, gender, or any other criteria than CD4 threshold.  Of individuals dropping out of the ART program, $50\%$ are eligible to re-enroll. See Eaton et al.~2012 for a detailed discussion of the interventions modeled in the study \cite{eaton_systematic_2012}.

\subsection{Timecourse of a Simulation}

\begin{figure*}[t]
\centering
\includegraphics[width=.9\textwidth]{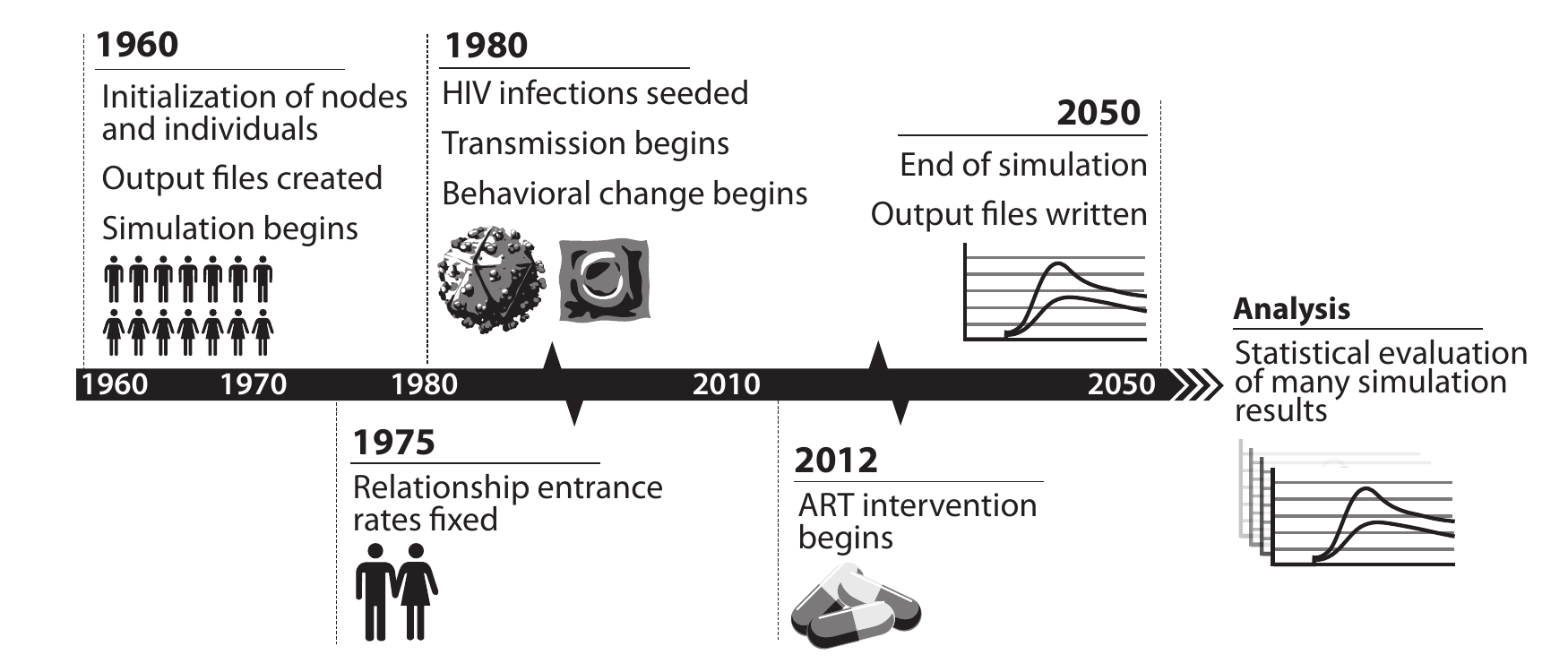}
\caption{The process of an EMOD-HIV simulation is depicted graphically. After initialization steps such as memory allocation and parsing of input files, the simulation begins in $1960$. Newly instantiated individuals with demographic properties dictated by the demographics module begin forming relationships at rates that are adjusted daily based on the MOPAD. After a burn-in period of $15$ years, the entrance rate adjustment is stopped, and the age-dependent entry rates into each PFA are fixed at their $1975$ rates for the rest of the simulation. In $1980$, HIV infection appears in $5\%$ of the population at random, with their time since infection sampled uniformly from their total untreated HIV prognosis. This high initial seeding of infection was used to offset the disproportionate number of low-risk infections created by this method of seeding the early epidemic. The intervention (ART distribution according to a variety of criteria and coverage levels) is applied instantaneously in $2012$, and the simulation continues until $2050$. Output files from numerous simulations are combined for statistical analysis in Matlab.}
\label{f:timecourse}
\end{figure*}

The timecourse of each simulation run is depicted graphically in Figure \ref{f:timecourse}.  Simulation begins in year $1960$ to allow ample time for relationships to ``burn-in.''  During the period between $1960$ and $1975$, the relationship formation rates for each gender and relationship type are updated daily using the relationship flow algorithm.  Adjustment of PFA entry rates is terminated in $1975$, and the rates are fixed at their $1975$ values for the remainder of the simulation.

Infections are seeded in year $1980$. The seeding process infects $5\%$ of the population according to the age and gender distribution reported in the $2008$ HSRC survey \cite{shisana_south_2010}.  This prevalence is higher than values generally accepted for $1980$ in South Africa; however, reproducing the historical trend of HIV prevalence would require a model that acknowledges the behavior change that has occurred from the $1980$'s until the present day as a result of increasing HIV awareness. These include reductions in commercial sex work and other high-risk activities, as well as increases in protective behavior such as condom usage, male circumcision, and serosorting-like partner choice. Since little data is available about behavior during the early HIV epidemic, we chose not to model these early-epidemic conditions, and therefore initialized our simulation at artificially high prevalence to account for the relatively lower-risk individuals who are initially infected in our model. 

The ART intervention is enabled in year $2012$.  Eligible individuals enroll in ART according to an exponential distribution having a mean of one year.  The ART intervention continues through the conclusion of the simulation in year $2050$.

\subsection{Model Calibration}

The EMOD-HIV $v0.7$ calibration process used data from the 2002, 2005, and 2008 HSRC surveys \cite{shisana_nelson_2002,shisana_south_2008,shisana_south_2010} in addition to antenatal prevalence data from $1990$ to $2009$ \cite{anc2010}. The parameters that were subject to the most careful calibration were those to which the epidemic trajectory was highly sensitive, but for which available data is insufficient to know their value with reasonable certainty. These include the absolute per-contact transmission rate, the rates of condom use by year in each relationship type, and the probability of concurrency in each relationship type.

As previously mentioned, the calibration process used to generate the results presented in Eaton et al.~$2012$ was rudimentary.  Parameter values were initialized to values reported in field studies, even when these values have high uncertainty and likelihood of survey bias.  Then, a coordinate descent with golden section search was used to incrementally adjust parameter values, one at a time, so as to minimize a squared error loss function.  This function penalized differences between the simulation output and HSRC prevalence data by age and gender, and ANC prevalence as compared to simulated females aged $25$ to $30$.  Parameters that are known with the lowest certainty were prioritized for adjustment.

This approach to calibration worked to a limited extent, but was unable to resolve co-varying parameters.  These parameters were adjusted by hand in order to minimize the same loss function.  Subsequent versions of the EMOD-HIV model will employ a sophisticated calibration process that uses Bayesian methods.

\section{Conclusions and future work}

The EMOD-HIV model is a work in progress.  Assumptions have been improved and numerous features have been added since the initial Stellenbosch Modelling Consortium meeting.  One area of improvement is model calibration.  As previously mentioned, the model was initially calibrated using a rudimentary algorithm that required much adjustment by hand.  We have since developed a Bayesian calibration tool based on Incremental Mixture Importance Sampling \cite{raftery2010estimating}.  Improved calibration will allow parameter uncertainty to be included in estimates in addition to finding appropriate values for key parameters.  
Future work will build upon the model described here.  Features such as increased detail in our representation of STIs and viral-load driven transmission are planned for the immediate future.  Additionally, the model will soon gain the ability to perform spatial simulation by using the spatial support features of the EMOD software implemented for all diseases.  Migration between geographic locations, representing communities, will be built using realistic migration patterns.  Finally, we plan to analyze a variety of interventions such as circumcision, ART treatment options, and vaccine candidates.

\section{Acknowledgments}
The authors thank Bill and Melinda Gates for their active support of this work and their sponsorship through the Global Good Fund. This work was performed at Intellectual Ventures Laboratories with help from colleagues, especially Grace Huynh, Christopher Lorton, Karima Nigmatulina, Gene Oates, and Michael Schnall-Levin. Helpful discussions with members of the HIV Modelling Consortium and the Bill and Melinda Gates Foundation, especially Timothy B.~Hallett, Geoffrey P.~Garnett, and Christine Rousseau are likewise gratefully acknowledged.

\bstctlcite{IEEEexample:BSTcontrol}

\bibliographystyle{IEEEtran}
\bibliography{EMODHIV}

\appendix

\onecolumn 

\section{Parameter Table}

\tablefirsthead{Parameter Table\\}
\tablehead{Parameter Table Continued\\}
\tabletail{}
\tablelasttail{}
\begin{supertabular*}{\textwidth}{p{12eM}ccp{16eM}}
\toprule
	{\bf Parameter}&	{\bf Value}&	{\bf Units}&	{\bf Comments}\\\\
\multicolumn{4}{c}{General}\\\midrule
	Initial population&	75000&	people&	Individuals uniformly weighted to represent full S.A.~population.\\
	Number of burn-in days&	7300&	days&	Days from 1960 to infection seeding in 1980.\\
	Duration of simulation&	32850&	days&	Days from 1960 to end in 2050.\\
	Initial prevalence fraction&	5&	percent&	Seeding prevalence, high for 1980 but gives good 2010 results.\\
	Circumcision probability&	42&	percent&	Nationally, $42\%$ of men aged 16-55 years report being circumcised.\\
	STI probability&	7&	percent&	Approximates fraction of STI (HSV-2) infected individuals that would be shedding.\\
	Minimum age for sexual interaction&	13&	years&	Less than $10\%$ have sex before 15 \cite{shisana_south_2010}.  Sex before 13 is excluded.\\
	Coital frequency&	0.33&	acts/day&	Self-reported coital frequency among discordant couples in Rakai, Uganda \cite{wawer_rates_2005}.\\\\
\multicolumn{4}{c}{Network}\\\midrule
	Male mean debut age&	16.5&	years&	Mean debut age for males reported as $16.3$ \cite{johnson_second_2010}.\\
	Female mean debut age&	15.5&	years&	Calibrated from mean debut of 17.2 reported \cite{johnson_second_2010}.\\
	Days between adding relationships&	60&	days&	Avoids excessive concurrent relationships, calibrated.\\
	Coital dilution with 2 partners&	75&	percent&\multirow{3}{16eM}{\vfill Relatively modest parameterization based on conflicting data \cite{sawers_hiv_2011,sawers_concurrent_2010,blower_sex_1993,reniers_polygyny_2010,josephson_does_2002,pebley_polygyny_1988}.}\\
	Coital dilution with 3 partners&	60&	percent&	\\
	Coital dilution with 4+ partners&	45&	percent&	\\
	Disable entrance rate adjustment on sim day&	5475&	days&	Rate adjustment ends five years before seeding.\\
	Transitory relationship formation rate&	0.00137&	formations/day&	\multirow{6}{16eM}{\vfill Exponentially distributed formation and Weibull distributed breakups result in reasonable durations, relationship prevalence, and cumulative lifetime number of relationships.  Adjusted during calibration.}\\
	Transitory relationship duration (mean)&	0.003044&	breakups/day	&\\
	Transitory relationship duration (shape)&	1.2&	dimensionless&\\
	Informal relationship formation rate&	0.001442&	formations/day&\\
	Informal relationship duration (mean)&	0.001522&	breakups/day	&\\
	Informal relationship duration (shape)&	2.0&	dimensionless	&\\
	Marital relationship formation rate&	9.13E-5&	formations/day&\\
	Marital relationship duration (mean)&	1.37E-4&	breakups/day	&\\
	Marital relationship duration (shape)&	1.5&	dimensionless&\\
	Transitory PFA processing interval&	1&	day&	\multirow{3}{16eM}{\vfill Processing intervals ensure sufficient partner diversity while avoiding excessive delays.}\\
	Informal PFA processing interval&	1&	day&	\\
	Marital PFA processing interval&	14&	days&	\\
	Maximum simultaneous transitories&	3&	relationships&	\multirow{3}{16eM}{\vfill Selected during calibration to arrive at a ~50 lifetime partners max while enabling concurrency.   Polygamy is not enabled.}\\
	Maximum simultaneous informals&	2&	relationships&\\
	Maximum simultaneous marriages&	1&	relationships&	\\
	Minimum age for marriage&	15&	years&	Marriage before 15 is unusual in South Africa \cite{johnson_second_2010}, as reflected in the marital MOPAD matrix.\\
	Extrarelational probability transitory&	80&	percent&	\multirow{3}{16eM}{\vfill Calibrated to get HSRC and ANC prevalence to align while avoiding too many lifetime partners.}\\
	Extrarelational probability informal&	30&	percent&	\\
	Extrarelational probability marriage&	10&	percent&\\\\
\multicolumn{4}{c}{Sexual Transmission}\\\midrule
Acute prob transmission per act&	0.034027&	prob/act&	Relative values adapted from analysis of transmission among discordant couples in Rakai, Uganda \cite{hollingsworth_hiv_1_2008}; absolute rates adjusted by 1.5 factor during calibration, and account for act frequency of 0.33 acts/day.  Values consistent with meta-analysis \cite{boily_heterosexual_2009}.\\
	Latent prob transmission per act&	0.001307&	prob/act&	\\
	AIDS prob transmission per act&	0.00937&	prob/act&	\\
	ART viral suppression risk multiplier&	0.04&	dimensionless&	Based on HPTN052 trial \cite{cohen_prevention_2011}\\
	Circumcision risk multiplier&	0.4&	dimensionless&	Trials in South Africa, Kenya, and Uganda \cite{auvert_randomized_2005,bailey_male_2007,gray_male_2007}\\
	Anal sex probability&	2.0&	percent&	$2.3\%$ in men and $1.2\%$ in women reported in US \cite{halperin1999heterosexual}.\\
	Anal risk multiplier&	10&	dimensionless&	$1.4\%$ per unprotected receptive act from a meta-analysis \cite{baggaley_hiv_2010,halperin_high, boily_heterosexual_2009}\\
	STI risk multiplier& 	4.0&	dimensionless&	2.58 for those with GUD in Rakai analysis \cite{gray_probability_2001}; 2-5 fold multiplier for HSV infection \cite{corey2004effects}.\\
	Condom risk multiplier&	0.1&	dimensionless&	$80\%$ reduction in incidence \cite{weller2002condom} and $78\%$ reduction found per act \cite{hughes2012determinants}\\\\
\multicolumn{4}{c}{Vertical Transmission}\\\midrule
	Usage of sdNVP&	45&	percent/birth&	Adherence of $57.1\%$ in Zimbabwe \cite{kuonza2010non}; $75\%$ adherence in S Africa \cite{peltzer2010antiretroviral}\\
	Mother to child transmission on sdNVP&	10&	percent/birth&	SWEN trial risk in single dose group $8.98\%$ at 6 months with single dose NVP; HIVNET 012 $11.8\%$ and $15.7\%$ at 6 week and 18 mo \cite{jackson2003intrapartum,guay_intrapartum_1999,martinson_transmission_2007}.\\
	Mother to child transmission no intervention&	30&	percent/birth&	$30\%$ (range 14-$48\%$) listed in \cite{msellati1995rates,newell_child_2004-1,orlando_cost-effectiveness_2010,de2000prevention}.\\
	Mother to child transmission on ART&	3&	percent/birth&	$2.9\%$ at 203 days (excludes infants born HIV+) \cite{chasela2010maternal}; $2.5\%$ at birth and $7.0\%$ at 24 mo in Kisumu breastfeeding study \cite{thomas2011triple}.  See also \cite{chibwesha_optimal_2011}.\\\\
\multicolumn{4}{c}{ART Intervention}\\\midrule
	Enable ART&	T or F&	&	Varies across intervention scenarios, false (F) for counterfactual.\\
	Year of ART initiation&	2012&&		Given by intervention scenario.\\
	Exponential ART dropout rate&	0.000263&&		Varies across intervention scenarios.\\
	Allow ART re-enrollment&	T&&		Re-enrollment is enabled.\\
	Probability of ART re-enrollment&	50&	percent&	Independent probabilities for each re-enrollment event.\\\\
\multicolumn{4}{c}{Condom Sigmoid	}\\\midrule
Condom low transitory&	1.3&	percent/act&	 	\multirow{12}{16eM}{\vfill Values calibrated based loosely on condom usage data reported in national surveys.}\\
	Condom high transitory&	64&	percent/act	&\\
	Condom mid year transitory&	1999&	 year	&\\
	Condom rate transitory&	0.3&	1/year	&\\
	Condom low informal&	1.3&	percent/act&\\	
	Condom high informal&	51&	percent/act	&\\
	Condom mid year informal&	1999&	 year	&\\
	Condom rate informal&	0.25&	1/year	&\\
	Condom low marriage&	6.25E-2&	percent/act	&\\
	Condom high marriage&	20&	percent/act&\\	
	Condom mid year marriage&	2002&	 year&\\	
	Condom rate marriage&	0.225&	1/year	&\\
\bottomrule
\end{supertabular*}


\end{document}